\documentclass[%
letterpaper,
reprint,
frontmatterverbose, 
preprintnumbers,
amsmath,amssymb,
aps,
pra,
showkeys,
floatfix,
]{revtex4-2}

\usepackage[per-mode = symbol]{siunitx}
\let\svqty\qty
\usepackage{physics}
\let\qty\svqty
\usepackage{mathtools}
\usepackage{xfrac}
\usepackage{graphicx}
\usepackage[caption=false,position=top]{subfig}
\usepackage{ragged2e}
\usepackage{dcolumn}
\usepackage{multirow}
\usepackage{bm}
\usepackage{hyperref}

\usepackage{lineno}

\makeatletter
\let\LN@align\align
\let\LN@endalign\endalign
\renewcommand{\align}{\linenomath\LN@align}
\renewcommand{\endalign}{\LN@endalign\endlinenomath}
\let\LN@gather\gather
\let\LN@endgather\endgather
\renewcommand{\gather}{\linenomath\LN@gather}
\renewcommand{\endgather}{\LN@endgather\endlinenomath}
\makeatother

\makeatletter
\def\@fnsymbol#1{\ensuremath{\ifcase#1\or \dagger\or *\or \ddagger\or
   \mathsection\or \mathparagraph\or \|\or **\or \dagger\dagger
   \or \ddagger\ddagger \else\@ctrerr\fi}}
\makeatother

\newcommand{\np}{\text{\tiny{NP}}}
\allowdisplaybreaks

\begin{document}


\title{Numerical modeling of the multi-stage Stern--Gerlach experiment by Frisch and Segr\`e using co-quantum dynamics via the Bloch equation}

\author{Kelvin Titimbo}
\thanks{These authors contributed equally.}
\affiliation{Caltech Optical Imaging Laboratory, Andrew and Peggy Cherng Department of Medical Engineering, Department of Electrical Engineering, California Institute of Technology, 1200 E. California Blvd., MC 138-78, Pasadena, CA 91125, USA}

\author{David C. Garrett}
\thanks{These authors contributed equally.}
\affiliation{Caltech Optical Imaging Laboratory, Andrew and Peggy Cherng Department of Medical Engineering, Department of Electrical Engineering, California Institute of Technology, 1200 E. California Blvd., MC 138-78, Pasadena, CA 91125, USA}

\author{S.~S\"uleyman Kahraman}
\thanks{These authors contributed equally.}
\affiliation{Caltech Optical Imaging Laboratory, Andrew and Peggy Cherng Department of Medical Engineering, Department of Electrical Engineering, California Institute of Technology, 1200 E. California Blvd., MC 138-78, Pasadena, CA 91125, USA}

\author{Zhe He}
\thanks{These authors contributed equally.}
\affiliation{Caltech Optical Imaging Laboratory, Andrew and Peggy Cherng Department of Medical Engineering, Department of Electrical Engineering, California Institute of Technology, 1200 E. California Blvd., MC 138-78, Pasadena, CA 91125, USA}

\author{Lihong V.~Wang}
\email[Email:]{lvw@caltech.edu}
\affiliation{Caltech Optical Imaging Laboratory, Andrew and Peggy Cherng Department of Medical Engineering, Department of Electrical Engineering, California Institute of Technology, 1200 E. California Blvd., MC 138-78, Pasadena, CA 91125, USA}

\date{\today}%

\begin{abstract}
We numerically study the spin flip in the Frisch--Segr\`e experiment, the first multi-stage Stern--Gerlach experiment, within the context of the novel co-quantum dynamics theory.
We model the middle stage responsible for spin rotation by sampling the atoms with the Monte Carlo method and solving the dynamics of the electron and nuclear magnetic moments numerically according to the Bloch equation. 
Our results show that, without using any fitting parameters, the co-quantum dynamics closely reproduces the experimental observation reported by Frisch and Segr\`e in 1933, which has so far lacked theoretical predictions.
\end{abstract}

\keywords{spin-flip transition, electron spin, Bloch equation, co-quantum dynamics.}
\maketitle

\section{Introduction}

The Stern--Gerlach experiment was a crucial benchmark for the early development of quantum mechanics \cite{gers1922,gers1924} and is still presented in introductory books as evidence of quantization and the existence of the electron spin angular momentum \cite{uhlg1925,est1975,braj1986,gre1998,mes2020}.
The so-called Stern--Gerlach apparatus (SG) has been used to illustrate the projection of the quantum wave function onto its eigenstates along the quantization axis, which is given by the direction of a strong inhomogeneous magnetic field.
Despite quantum mechanics being fundamental for explaining physical phenomena in almost any branch of physics, there is no generally accepted theory for how the wave function collapses \cite{wig1963,buslm1996, sch2005a,basls2013}.

Recently, a novel theory denoted co-quantum dynamics (CQD) has been proposed to describe the evolution and collapse of electron spins in alkali atoms interacting with an external magnetic field, $\vb{B}$ \cite{wan2022,wan2022a}. 
In CQD, the electron magnetic moment $\vb{\bm{\mu}}_{e}$ is termed the principal quantum, whereas the nuclear magnetic moment $\vb{\bm{\mu}}_{n}$ is termed the co-quantum.
The evolution of both quanta is modeled by the Bloch equation, and the collapse of the principal quantum is treated by adding an induction term.
In addition to $\vb{B}$, the magnetic field $\vb{B}_{n}$ generated by the nuclear magnetic moment also acts upon $\vb{\bm{\mu}}_{e}$; similarly, the magnetic field $\vb{B}_{e}$ generated by the electron magnetic moment  acts upon $\vb{\bm{\mu}}_{n}$. 

We apply CQD to model the multi-stage Stern--Gerlach experiment conducted by R.~Frisch and E.~Segr\`e \cite{fris1933,fris1933a}.
Their apparatus is composed of two Stern--Gerlach stages and an intervening stage referred to as the inner rotation chamber. 
The middle stage is characterized by the presence of a weak but rapidly varying magnetic field that rotates the electron magnetic moment. 
Despite attempts by E.~Majorana \cite{maj1932} and I.I.~Rabi \cite{rab1936,rab1937,blor1945}, theoretical descriptions of this experiment deviate from the Frisch--Segr\`e experimental observation \cite{wan2022,schsl2016}.
To date, this divergence remains unresolved by existing theories \cite{schsl2016,fris2021,kahth2022}.
However, a closed-form approximation of CQD matches the experimental data with a high coefficient of determination without the use of fitting parameters \cite{wan2022}.
The approximation models the dynamics by means of the Schr\"odinger equation coupled with the CQD concept; this approach has been corroborated numerically \cite{hetk2022}.
Here, we validate CQD by numerically solving the Bloch equation instead to estimate the fraction of spin flip in the Frisch--Segr\`e experiment and achieve a high coefficient of determination.

The manuscript is organized as follows. 
The experimental setup used by Frisch and Segr\`e is described in Section \ref{physys}. 
The CQD theory for the experiment is described  in Section \ref{theo}. 
First, we introduce the equations of motion for the electron and nuclear magnetic moments. 
Second, we describe all stages in the Frisch--Segr\`e experiment with emphasis on the intermediate stage.
We compare our numerical results with both the experimental observation and the analytical solution reported previously \cite{wan2022}.
In Section \ref{conclu}, we make final remarks about our findings.
In the Appendices, we mathematically derive the approximated external magnetic field (Appendix \ref{app:quadru}) and summarize the closed-form analytical formula obtained using CQD (Appendix \ref{app:CQD}).

\section{Experiment}\label{physys}

\begin{figure}
    \centering
    \includegraphics[width=0.95\linewidth]{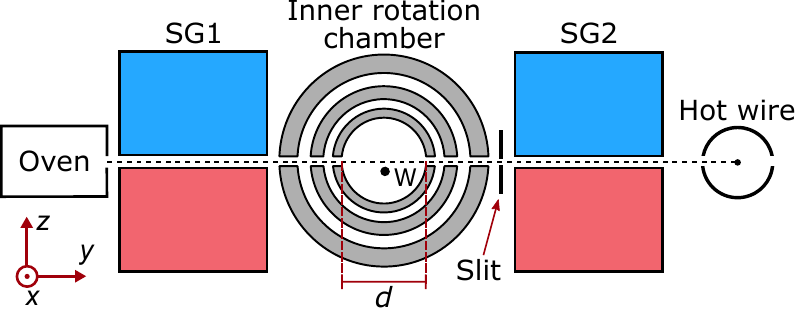} 
    \caption{Schematic of the Frisch--Segr\`e apparatus as in \cite{fris1933,fris1933a}. The atomic beam from the oven is sent to SG1, whose inhomogeneous magnetic field points along the $z$-axis.
    The blue color indicates the magnetic south pole, whereas the red color indicates the magnetic north pole.
    Then, the atomic beam enters the magnetically shielded inner rotation chamber containing a remnant homogeneous magnetic field plus a magnetic field generated by the electric current $I_{w}$ flowing through a wire $W$ along the $-x$-direction. A slit post-selects one of the initially aligned branches of the electron magnetic moment. Finally, the atoms travel through SG2, and the fraction of spin flip is measured using a hot wire.}
    \label{fig:setup}
\end{figure}

A schematic of the multi-stage Stern--Gerlach apparatus, first constructed by T.E.~Phipps and O.~Stern \cite{phis1932} and later improved by Frisch and Segr\`e \cite{fris1933,fris1933a}, is shown in Fig.~\ref{fig:setup}.
The experimental setup combines two magnet pairs, SG1 and SG2, which each act as a standard Stern--Gerlach apparatus \cite{gers1924,sty2000, mes2020}. 
Between them, a magnetically shielded inner rotation chamber contains a homogeneous remnant magnetic field and a cylindrically symmetric magnetic field generated by a current-carrying wire.
The aim of the experiment is to observe spin flips in ground-state alkali atoms due to non-adiabatic rotations in the middle stage.

Alkali atoms are appealing for this experiment since their ground state has a closed shell with only one valence electron having an orbital angular momentum $\vb{L} =0$.
As a consequence, the total angular momentum for the electron of the atom equals the spin $\vb{S}$, with $S=\sfrac{1}{2}$, and the spin{\textendash}orbit coupling vanishes.
However, the electron{\textendash}nuclear spin interaction, also called the hyperfine interaction, arises from the coupling between the nuclear magnetic moment $\vb{\bm{\mu}}_{n}$ and the magnetic field generated by the electron magnetic moment $\vb{\bm{\mu}}_{e}$, or equivalently vice versa.
The coupling strength depends on the relative orientations of the magnetic moments, which are related to the spin angular momenta as
\begin{align}\label{eq:magneticmoments}
    \vb{\bm{\mu}}_{e}&=\gamma_{e} \vb{S}, & \vb{\bm{\mu}}_{n}&=\gamma_{n} \vb{I},
\end{align}
where $\gamma_{e}$ is the electron gyromagnetic ratio, $\gamma_{n}$ is the nuclear gyromagnetic ratio, and $\vb{I}$ is the nuclear spin. 

\begin{figure}
    \centering
    \includegraphics[width=0.95\linewidth]{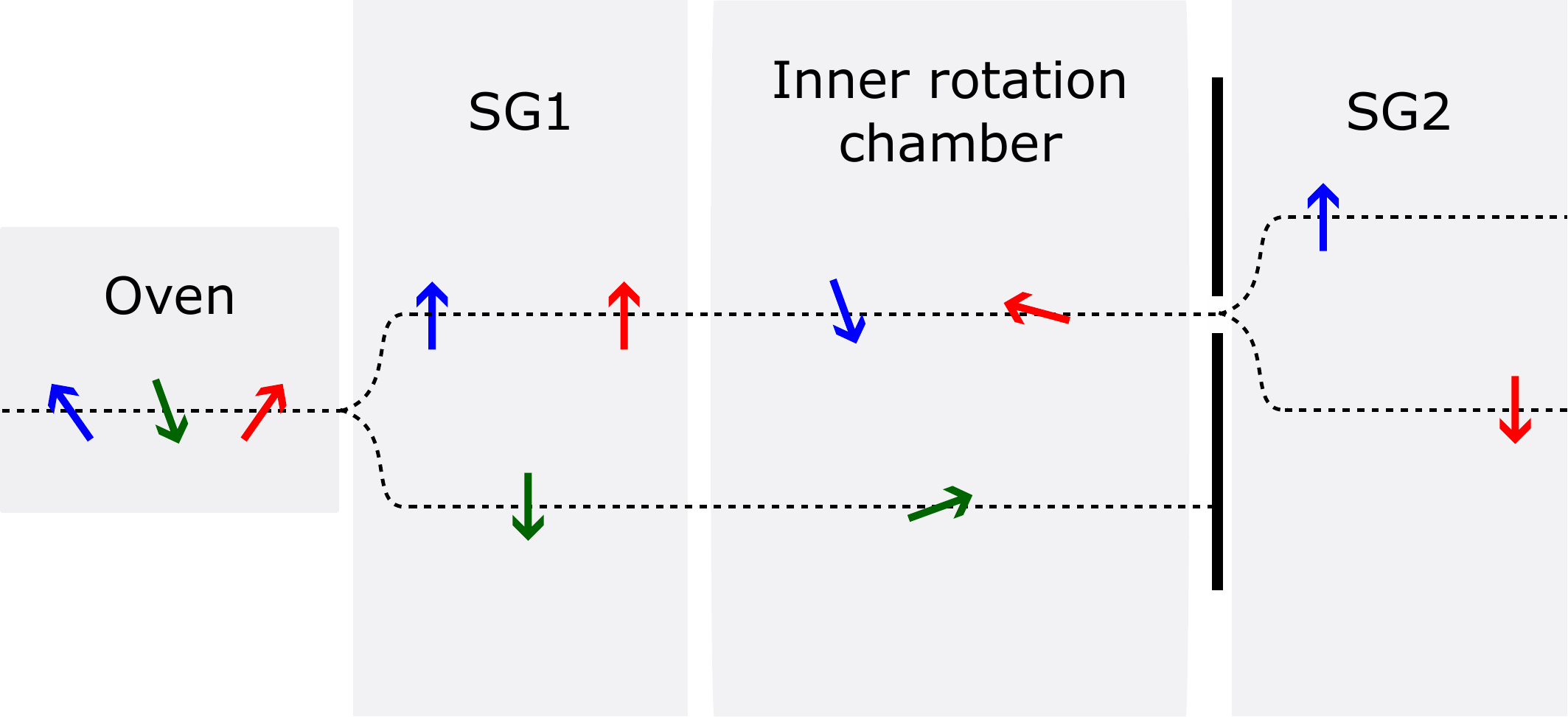}
    \caption{Illustration of three representative electron magnetic moments of atoms traversing the Frisch--Segr\`e apparatus. At the oven, the electron magnetic moments are randomly oriented, SG1 aligns them to $\uparrow$ and $\downarrow$ with respect to the orientation of the magnetic field. Both branches enter the inner rotation chamber, where the varying magnetic field rotates the electron magnetic moments. The upper branch is selected using a slit, and the atomic beam is sent to SG2 where the electron magnetic moments are again aligned. Finally, both branches are measured to quantify the fraction of spin flip. }
    \label{fig:mm}
\end{figure}

A beam of alkali atoms is easily generated since alkali metals have sufficient vapor pressure at temperatures of only a few hundred degrees Celsius \cite{fior1926,alcih1984}.
Let us consider the beam of atoms emerging from an oven and propagating along the $y$ direction as shown in Fig.~\ref{fig:setup}.
The atomic beam first enters SG1, the strong magnetic field ($\vb{B}_{0} \sim\SI{0.5}{\tesla}$) along the $z$-axis defines the quantization axis. 
The initially isotropically oriented electron magnetic moments are quickly aligned parallel $\uparrow$ or antiparallel $\downarrow$ to the magnetic field direction. 
The gradient of the magnetic field deflects the atoms into two branches.
Despite no branch being physically selected before the inner rotation chamber, in the following description, we track only the branch with magnetic moments parallel to the field \cite{fris1933a}. 
Figure \ref{fig:mm} depicts the general behavior of electron magnetic moments through the experimental setup.

\begin{figure}[t]
\captionsetup[subfloat]{justification=Justified, singlelinecheck=false}
\centering
    \subfloat[\label{fig:Ba}]{\includegraphics[width=0.82\columnwidth]{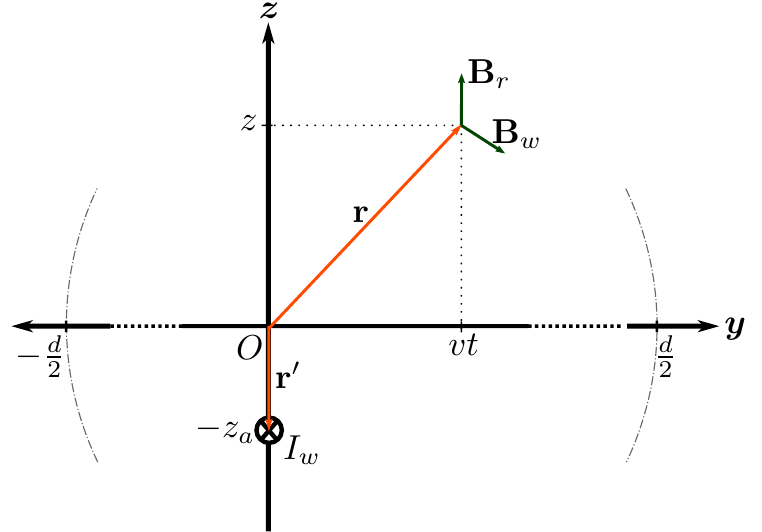}} 
    
    \subfloat[\label{fig:Bb}]{\includegraphics[width=0.82\columnwidth]{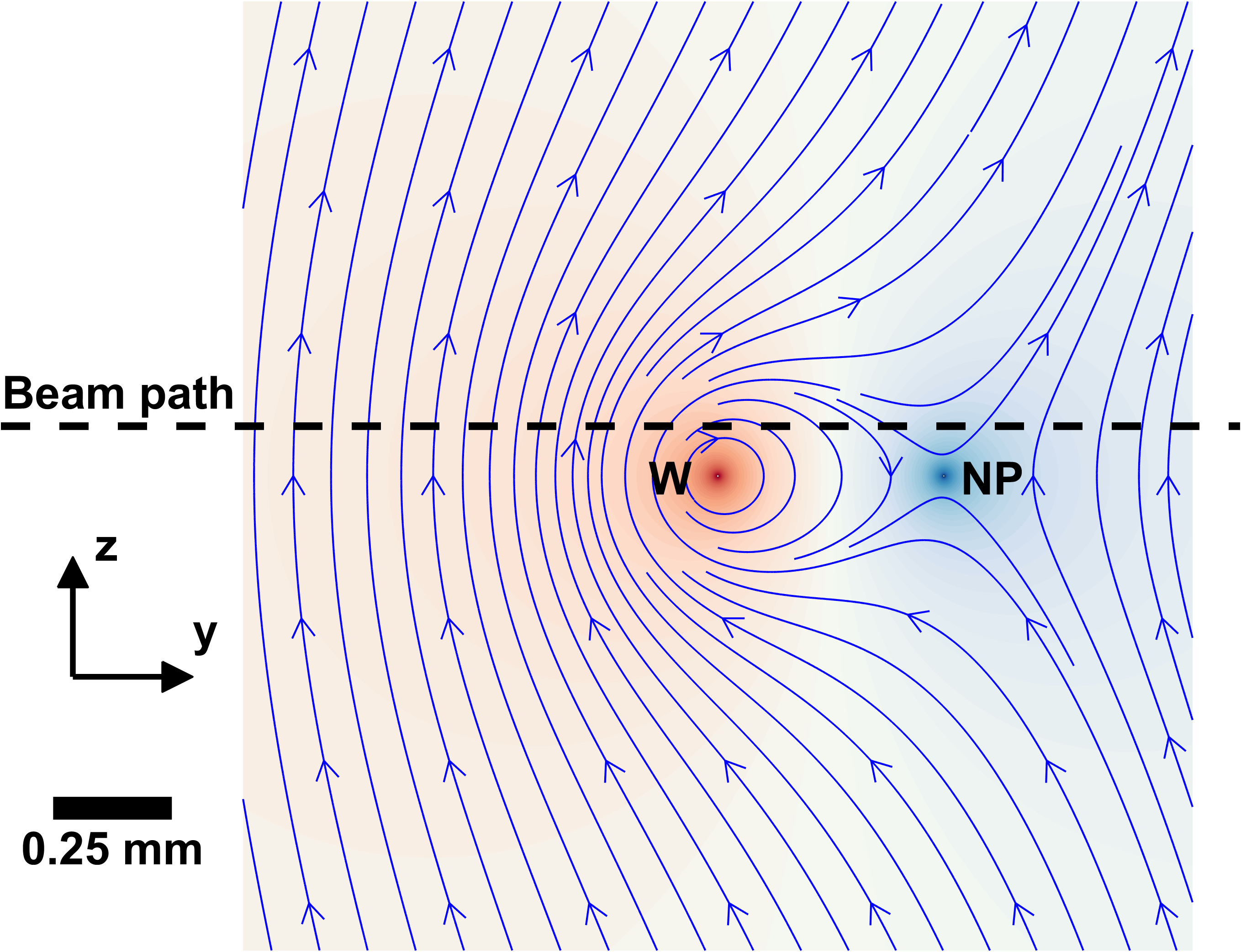}} 
\caption{Magnetic field inside the inner rotation chamber. The beam propagates along the $y$-axis at constant speed $v$. The top panel shows the Cartesian reference frame, the position of the current-carrying wire (not drawn to scale), and the contributions from the remnant field $\vb{B}_{r}$ and the wire field $\vb{B}_{w}$ to the total magnetic field at position $\vb{r}$. The bottom panel shows the field lines of the total magnetic field with a null point (NP) below the beam path. In this example from the Frisch--Segr\`e experiment \cite{fris1933}, ${B}_{r} = \SI{42}{\micro\tesla}$, $I_{w}=\SI{0.02}{\ampere}$, and $z_{a}=\SI{105}{\micro\metre}$.}
\label{fig:Bfield}
\end{figure}

Next, the split beam is sent into an inner rotation chamber, which has an innermost diameter $d$.
Despite the shielding from the fringe fields of SG1 and SG2, in the innermost chamber a weak remnant magnetic field $\vb{B}_{r}$ persists, predominantly aligned along the $z$-axis.
A wire with electric current $I_{w}$, flowing along the $-x$ direction, is placed at a vertical distance $z_{a}$ below the selected atomic beam path.
Figure \ref{fig:Ba} shows the main elements and the coordinate system inside the chamber.
For an infinitely long straight wire, the magnetic field generated by the current, $\forall \, y,z$ in the chamber, is given by
\begin{equation} \label{eq:Bw}
    \vb{B}_{w} = \frac{\mu_{0}\, I_{w}}{2\pi \left[ (z+z_{a})^{2} + y^{2}  \right]} \left[  (z+z_{a})\, \vu{y} - y\, \vu{z} \right] \ ,
\end{equation}
where $\mu_{0}$ denotes the vacuum permeability. 
Therefore, the total field in the innermost shielded region is $\vb{B} = \vb{B}_{w} + \vb{B}_{r}$. 
As illustrated in Fig.~\ref{fig:Bb}, the magnetic field cancels at the null point (NP), at position $\vb{r}_{\np}= y_{\np} \vu{y} - z_{a} \vu{z}$, with 
\begin{equation}\label{eq:ynp}
    y_{\np} = \frac{\mu_{0} \, I_{w}}{2\pi B_{r}}  \ .
\end{equation}

In the region near the NP, a first-order Taylor expansion around the point $(y_{\np},-z_{a})$ approximates the magnetic field as a quadrupole (see Appendix \ref{app:quadru} for details):
\begin{equation}\label{eq:Bq}
    \vb{B} \xrightarrow[\text{vicinity}]{\np} \vb{B}_{q} =  \dfrac{2\pi B_{r}^{2}}{\mu_{0} \, I_{w}} \biggl[  (z+z_{a}) \vu{y} + (y-y_{\np}) \vu{z} \biggr] .
\end{equation}

After the atom interacts with the field inside the inner rotation chamber, the upper branch is selected using a slit while the lower branch is blocked, as illustrated in Fig.~\ref{fig:mm}. 
Frisch and Segr\`e claimed that placing the slit immediately after instead of before the inner rotation chamber improved the beam quality \cite{fris1933}.

Finally, the selected atomic branch enters SG2.
This second Stern--Gerlach stage again aligns the electron magnetic moment of each atom in parallel $\uparrow$ and antiparallel $\downarrow$ directions with respect to the strong magnetic field $\vb{B}_{0}$ along the $+z$-axis.
A hot wire records the final distribution of the atoms, which is used to quantify the fraction of spin flip of the electron magnetic moment resulting from the inner rotation chamber.
The experiment is repeated with varying wire currents $I_{w}$, producing a fraction of spin flip as a function of the current.

\section{Theory}\label{theo}

Here, we use CQD \cite{wan2022a,wan2022} to describe the multi-stage Stern--Gerlach experiment by Frisch and Segr\`e.
The formalism provides a mechanism of collapse of the electron spin in atoms interacting with an external magnetic field based on the combined dynamics of the electron and nuclear magnetic moments \eqref{eq:magneticmoments}.
The Bloch equation governs the evolution of the magnetic moments in the inner rotation chamber.

\subsection{Magnetic moment evolution}

The evolution of the magnetic moments in an external magnetic field are described by the Bloch equation \cite{blo1946}:
\begin{equation} \label{eq:bloch}
    \dv{\vu{\bm{\mu}}}{t} = \gamma \, \vu{\bm{\mu}} \cross \vb{B} \, ,
\end{equation}
where $\vu{\bm{\mu}}$ denotes the unit vector of the magnetic moment, $\gamma$ is the gyromagnetic ratio, and $\vb{B}$ is the magnetic flux density.
This equation governs an undamped precession of the magnetic moment about the magnetic field $\vb{B}$.
While usually considered a classical formalism, the Bloch equation has been recently shown to yield the space-independent von Neumann equation \cite{wan2022a}. 

For the principal quantum $\vb{\bm{\mu}}_{e}$ and the co-quantum $\vb{\bm{\mu}}_{n}$, the Bloch equation \eqref{eq:bloch} becomes
\begin{align}
    \label{eq:beme} \dv{\vu{\bm{\mu}}_{e}}{t} &= \gamma_{e} \, \vu{\bm{\mu}}_{e}\cross(\vb{B}+\vb{B}_{n}) \, ,\\
    \label{eq:bemu} \dv{\vu{\bm{\mu}}_{n}}{t} &= \gamma_{n} \,  \vu*{\bm{\mu}}_{n}\cross(\vb{B}+\vb{B}_{e}) \, ,
\end{align}
where $\gamma_{e}$ and $\gamma_{n}$ denote the gyromagnetic ratios of the electron and the nucleus, respectively, and $\vb{B}$ the external magnetic field.
CQD introduces the $\vb{\bm{\mu}}_{e}\text{\textendash}\vb{\bm{\mu}}_{n}$ interaction via each other's torque-averaged magnetic field, given by \cite{wan2022}
\begin{align}
    \label{eq:Be} \vb{B}_{e} &= \dfrac{5\mu_{0} \, \mu_{e}}{16\pi R^{3}} \vu{\bm{\mu}}_{e} = B_{e} \, \vu{\bm{\mu}}_{e} \, , \\
    \label{eq:Bn} \vb{B}_{n} &= \dfrac{5\mu_{0} \, \mu_{n}}{16\pi R^{3}} \vu{\bm{\mu}}_{n} = B_{n} \, \vu{\bm{\mu}}_{n} \, ,
\end{align}
where $R$ is the van der Waals radius of the atom, and $\mu_{e}$ and $\mu_{n}$ are the magnitudes of the magnetic moments for the electron and the nucleus, respectively.

The orientations of the magnetic moments in $\mathbb{R}^{3}$ are conveniently described using spherical coordinates.
With the polar and azimuthal angles, $\theta_{e}$ and $\phi_{e}$ for $\vb{\bm{\mu}}_{e}$ and $\theta_{n}$ and $\phi_{n}$ for $\vb{\bm{\mu}}_{n}$, we write the unit vectors as
\begin{equation}
    \label{eq:mue_sphe} \vu{\bm{\mu}}_{e} = \pmqty{\sin(\theta_{e})\cos(\phi_{e}) \\ \sin(\theta_{e})\sin(\phi_{e}) \\ \cos(\theta_{e})} ,
\end{equation}
\begin{equation}
    \label{eq:mun_sphe} \vu{\bm{\mu}}_{n} = \pmqty{\sin(\theta_{n})\cos(\phi_{n}) \\ \sin(\theta_{n})\sin(\phi_{n}) \\ \cos(\theta_{n})} .
\end{equation}
Meanwhile, the external field is written as
\begin{equation} \label{eq:Bvector}
    \vb{B} = \pmqty{B_{x}\\B_{y}\\B_{z}} .
\end{equation}

Finally, substituting \eqref{eq:Be}\textendash\eqref{eq:Bvector} into \eqref{eq:beme} and \eqref{eq:bemu}, we obtain a set of differential equations determining the evolution of the unit vectors of $\vb{\bm{\mu}}_{e}$ and $\vb{\bm{\mu}}_{n}$:
\begin{widetext}
\begin{subequations}\label{eq:fullbloch}
\begin{align}
    \dot{\theta}_{e} &= -\gamma_{e} \left[ B_{y} \cos(\phi_{e}) - B_{x}\sin(\phi_{e}) + B_{n} \sin(\theta_{n})\sin(\phi_{n}-\phi_{e})\right] \ , \\
    \dot{\phi}_{e} &= -\gamma_{e} \left[ B_{z} + B_{n} \cos(\theta_{n}) - \cot(\theta_{e})\left[ B_{x}\cos(\phi_{e}) + B_{y}\sin(\phi_{e}) + B_{n}\sin(\theta_{n})\cos(\phi_{e}-\phi_{n})  \right]  \right] \ , \\
    \dot{\theta}_{n} &= -\gamma_{n} \left[ B_{y} \cos(\phi_{n}) - B_{x}\sin(\phi_{n}) + B_{e} \sin(\theta_{e})\sin(\phi_{e}-\phi_{n})\right] \ , \\
    \dot{\phi}_{n} &= -\gamma_{n} \left[ B_{z} + B_{e} \cos(\theta_{e}) - \cot(\theta_{n})\left[ B_{x}\cos(\phi_{n}) + B_{y}\sin(\phi_{n}) + B_{e}\sin(\theta_{e})\cos(\phi_{e}-\phi_{n})  \right]  \right] \ .
\end{align}   
\end{subequations}
\end{widetext}

Given the dynamical equations for the electron and nuclear magnetic moments of a single atom, we model their evolution in the inner rotation chamber of the Frisch--Segr\`e experiment.

\subsection{Multi-stage Stern--Gerlach experiment by Frisch and Segr\`e}

In the Frisch--Segr\`e experiment, neutral potassium-39 atoms ($^{39}\mathrm{K}$) with an electron configuration of $[\mathrm{Ar}]\: 4\mathrm{s}^1$ were used.
The valence electron in its ground state is fully specified in the Russell-Saunders notation as $4\, ^{2}\mathrm{S}_{\sfrac{1}{2}}$.
Table \ref{tab:table1} lists the values of the van der Waals radius, spins, magnetic moments, and gyromagnetic ratios for the $^{39}\mathrm{K}$ atom. 
Accordingly, we compute the magnitudes of the magnetic fields generated by the electron and nuclear magnetic moments from \eqref{eq:Be} and \eqref{eq:Bn}:
\begin{subequations}
\begin{align}
    B_{e} &= \SI{55.80626722(2)}{\milli\tesla}, \\
    \label{eq:bn_value} B_{n} &= \SI{11.8842(3)}{\micro\tesla} \ .
\end{align}    
\end{subequations}

\begin{table}[b]
\caption{\label{tab:table1} Values for potassium-39 \cite{ariiv1977,antkk2012,sto2019,tiemn2021,rum2022}.}
\begin{ruledtabular}
\begin{tabular}{p{1mm}p{5mm}cc}
 \multicolumn{4}{c}{$^{39}\mathrm{K}$}\\
 & & Property & Value \\ \hline
 \multicolumn{2}{c}{Atom} & $R$ & $\phantom{\Bigl[}\SI{275}{\pico\meter}$ \\ \colrule 
  \parbox[t]{3mm}{\multirow{3}{*}{\rotatebox[origin=c]{90}{Valence \, }}} & \parbox[t]{2mm}{\multirow{3}{*}{\rotatebox[origin=c]{90}{electron \ }}} & $S$ & $\phantom{\Bigl[} \frac{1}{2} \phantom{\Bigl]}$    \\
 & & $\mu_{e}$ & $\phantom{\Bigl[} \SI{9.2847677043(28)e-24}{\joule\per\tesla}$\\
 & & $\gamma_{e}$ & $\phantom{\Bigl[} \SI{-1.76085963023(53)e11}{\radian\per\second\per\tesla}$ \\  \hline
 \multicolumn{2}{c}{\parbox[c]{3mm}{\multirow{3}{*}{\rotatebox[origin=c]{90}{Nucleus \: }}}} & $I$ & $\phantom{\Bigl[} \frac{3}{2} \phantom{\Bigl]} $ \\
 & & $\mu_{n}$ & $\phantom{\Bigl[} \SI{1.97723(4)e-27}{\joule\per\tesla}$ \\
 & & $\gamma_{n}$ & $\phantom{\Bigl[} \SI{1.2500612(3)e7}{\radian\per\second\per\tesla}$ \\
\end{tabular}
\end{ruledtabular}
\end{table}

\subsubsection{Oven}

First, heating $^{39}\mathrm{K}$ atoms in an oven with a small opening produces an effusive beam of non-interacting thermal atoms.
We define the $y$-axis along the beam as shown in Fig.~\ref{fig:setup}.

The initial orientations of both the electron and nuclear magnetic moments of a single traveling atom are isotropically distributed.
Therefore, their angular probability density functions are respectively
\begin{align*}
    p_{e,\mathrm{oven}}(\theta_{e,\mathrm{oven}},\phi_{e,\mathrm{oven}}) &= \frac{1}{4\pi} ,  \\
    p_{n,\mathrm{oven}}(\theta_{n,\mathrm{oven}},\phi_{n,\mathrm{oven}}) &= \frac{1}{4\pi} \ .
\end{align*}
To implement a Monte Carlo simulation, the polar and azimuthal angles are numerically generated from independent realizations of
\begin{align}
    \label{eq:eiso} \theta_{e,\mathrm{oven}} &= 2\arcsin(\sqrt{\zeta_{1}}),  & \phi_{e,\mathrm{oven}} &= 2\pi \zeta_{2}, \\
    \label{eq:niso} \theta_{n,\mathrm{oven}} &= 2\arcsin(\sqrt{\zeta_{3}}),  & \phi_{n,\mathrm{oven}} &= 2\pi \zeta_{4},
\end{align}
with $\zeta_{j}$, $j=1,\dots, 4$, being random numbers uniformly sampled from zero to one.

\subsubsection{Stern--Gerlach apparatus 1}

Emerging from the oven, the atoms enter a Stern--Gerlach apparatus consisting of an inhomogeneous magnetic field with a strong gradient along the $z$-axis, as illustrated in Fig.~\ref{fig:setup}.
Usually, the amplitude of the magnetic field is $B_{0} \sim \SI{0.5}{\tesla}$; thus, $B_{0} > B_{e} \gg B_{n}$.
For such a strong field, $\vb{\bm{\mu}}_{e}$ and $\vb{\bm{\mu}}_{n}$ couple more strongly to the external magnetic field than to each other, and can be considered as independently precessing with Larmor frequencies $\omega_{e}$ and $\omega_{n}$, for electron and nuclear magnetic moments, respectively, about the external field direction.
From the values in Table \ref{tab:table1}, the ratio between the nuclear and electron Larmor frequencies $\sfrac{\omega_{n}}{\omega_{e}} \approx \num{7e-5}$.
In consequence, while the flight time guarantees the collapse of the electron magnetic moment, it is too short for the nuclear magnetic moment to collapse \cite{wan2022}.

Therefore, the electron magnetic moments of the single atoms, initially isotropically distributed, split into two branches with well-defined opposite alignments along the $z$-axis.
We characterize the orientations of the aligned $\vb{\bm{\mu}}_{e}$ by setting $\theta_{e,0}= 0$ and $\theta_{e,0}=\pi$ in \eqref{eq:mue_sphe}, respectively, as illustrated in Fig.~\ref{fig:mm}.
Meanwhile, since $\vb{\bm{\mu}}_{n}$ does not collapse, CQD assumes that $\theta_{n}$ does not vary significantly during the flight.
CQD shows that at the end of the SG1 stage, the $\vb{\bm{\mu}}_{n}$ orientation in each of the branches is redistributed into an anisotropic probability density function \cite{wan2022}:
\begin{equation}\label{eq:HS}
    p_{n}(\theta_{n,0},\phi_{n,0}) =
    \begin{cases}
        \frac{1}{4\pi}\left( 1 - \cos(\theta_{n,0}) \right) & \text{if  } \theta_{e,0} = 0 , \\
        \frac{1}{4\pi}\left( 1 + \cos(\theta_{n,0}) \right) & \text{if  } \theta_{e,0} = \pi .
    \end{cases}
\end{equation}

\subsubsection{Inner rotation chamber}

The inner rotation chamber consists of three hollow iron shielding spheres. Without shielding, the fringe field was $\sim \SI{0.4}{\tesla}$; with shielding, the measured remnant field along the $+z$-axis was reported to be $B_{r}=\SI{42}{\micro\tesla}$ \cite{fris1933}.
Inside the chamber, an electric current $I_{w}$ flowing through a wire produces a spatially varying magnetic field \eqref{eq:Bw}, resulting in the total field shown in Fig.~\ref{fig:Bb}.

We select the branch of the atomic beam with $\vb{\bm{\mu}}_{e}$  aligned to the $+z$-axis according to the original experiment \cite{fris1933a}, \textit{i.e.,} $\theta_{e,0}=0$, as shown in Fig.~\ref{fig:mm}.
Therefore, the probability density function \eqref{eq:HS} becomes 
\begin{equation}\label{eq:hsd}
    p_{n}(\theta_{n,0},\phi_{n,0})=\frac{1}{4\pi}\left( 1 - \cos(\theta_{n,0}) \right) .
\end{equation}
To continue the Monte Carlo simulation, the polar and azimuthal angles $(\theta_{n,0},\phi_{n,0})$ are sampled from the above anisotropic probability density function:
\begin{align}
    \label{eq:hssamp} \theta_{n,0} &= 2\arcsin( \zeta_{1}^{1/4} ), & \phi_{n,0} &= 2\pi \, \zeta_{2} ,
\end{align}
where $\zeta_{1}$ and $\zeta_{2}$ are two independent random numbers sampled uniformly from zero to one.

Each atom in the beam propagates with velocity $\vb{v}=v \, \vu{y}$, and we approximate the motion to be rectilinear and uniform such that its position on the propagation axis is $y = v t $.
The atom is simulated from $-\sfrac{d}{2}$ to $\sfrac{d}{2}$ as shown in Fig.~\ref{fig:Ba}.
Here, $v=\SI{800}{\metre\per\second}$ and $d=\SI{16.3}{\milli\metre}$ \cite{fris1933}.
Therefore, the time in the inner rotation chamber $t \in \left[ \SI{-10.2}{\micro\second}, \SI{10.2}{\micro\second} \right]$.
The total magnetic field along the $y$-axis varies according to
\begin{equation}\label{eq:B}
    \vb{B} = \frac{\mu_{0} I_{w}}{2\pi \left[ z_{a}^{2} + (vt)^{2}  \right]} \left[  z_{a}\, \vu{y} - v t\, \vu{z} \right] + B_{r} \, \vu{z} \ .
\end{equation}
The wire is positioned along the $x$-axis at a distance $z_{a}=\SI{105}{\micro\meter}$ below the beam path.
The electric current $I_{w}$ varies from $\SI{0.01}{\ampere}$ to $\SI{0.5}{\ampere}$ \cite{fris1933,fris1933a}.

It is known that a time-dependent magnetic field can rotate the magnetic moment of the electron \cite{maj1932,blor1945,kofis2022,ivasn2023}.
To determine whether the process is adiabatic or non-adiabatic, we compute the so-called adiabaticity parameter, defined as the ratio between the absolute values of the Larmor frequency $\omega_{e}$ and the rotational speed of the field in the $yz$ plane $\Omega_{B}$ \cite{maj1932}:
\begin{equation}
    k = \abs{\frac{\omega_{e}}{\Omega_{B}}} = \abs{\dfrac{\gamma_{e} B}{ \dv{t} \! \left[ \arctan\left( {B_{z}}/{B_{y}}\right) \right] } }  \ .
\end{equation}
If $\omega_{e} \gg \Omega_{B} $ ($k\gg 1$), $\vb{\bm{\mu}}_{e}$ is able to follow the changing orientation of the magnetic field, and the rotation is said to be adiabatic. 
Otherwise, $\vb{\bm{\mu}}_{e}$ may be unable to follow the changing orientation of the magnetic field; if so, the rotation is said to be non-adiabatic.

\begin{figure}
    \centering
    \includegraphics[width=0.97\linewidth]{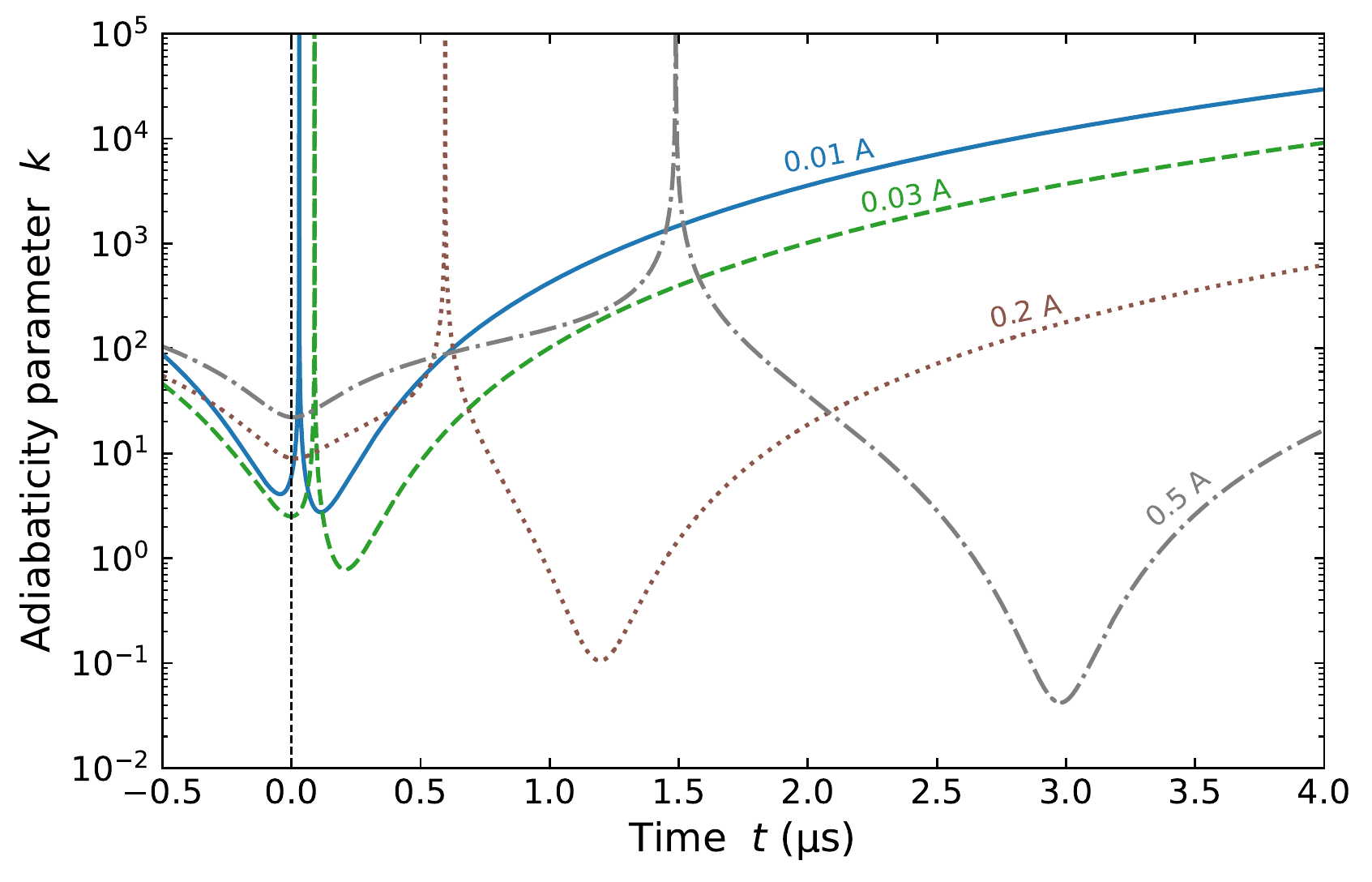} 
    \caption{Adiabaticity parameter $k$ along the beam path for various electric currents. Adiabatic rotation of the electron magnetic moment occurs around the peak ($k \gg 1$), whereas non-adiabatic rotation happens around the trough (near the null point $t_{\np}$ of the magnetic field). The vertical black dashed line indicates the time when the atom is above the wire ($t=0$).}
    \label{fig:kterm}
\end{figure}

From \eqref{eq:B}, the adiabaticity parameter becomes
\begin{multline} \label{eq:kw}
    k = \Biggl| \dfrac{\gamma_{e} \mu_{0} I_{w}}{2\pi z_{a} v} \dfrac{\sqrt{(vt)^{2} + z_{a}^{2}}}{\left(1-\dfrac{4\pi B_{r} v t}{\mu_{0} I_{w}} \right)} \\
    \times \left[ 1 - \dfrac{4\pi B_{r} v t}{\mu_{0} I_{w}} + \left( \dfrac{2\pi B_{r}}{\mu_{0} I_{w}} \right)^{2} \left( (v t)^{2} + z_{a}^{2} \right) \right]^{3/2} \Biggr| .
\end{multline}
Figure \ref{fig:kterm} shows the variation of $k$ around the center of the cavity computed from \eqref{eq:kw} for a set of representative electric currents $I_{w}$.

We note that in the vicinity of the wire position ($t=0$) and before the atoms reach the null point at time
\begin{equation}\label{eq:tnp}
    t_{\np} = \frac{y_{\np}}{v}=\frac{\mu_{0} \, I_{w}}{2\pi v B_{r}} ,
\end{equation}
the adiabaticity parameter is much greater than unity at the peaks.
Accordingly, as the $z$ component of the magnetic field changes its orientation (see Fig.~\ref{fig:Bfield}), the electron magnetic moment follows the varying $\vb{B}$-field adiabatically.
Therefore, the electron magnetic moment flips adiabatically, $\theta_{e,0}: 0 \mapsto \pi$, in this region. 
In comparison, Majorana stated that the flip is due to the orientation reversal of the quadrupole field along the flight path \cite{maj1932}.
For simplicity, we ignore the contribution from the nucleus \eqref{eq:bn_value} to $k$ because its much weaker field does not significantly affect the magnitudes of the adiabatic peaks in Fig.~\ref{fig:kterm}

Then, the electron magnetic moment is rotated non-adiabatically around the corresponding trough in Fig.~\ref{fig:kterm}, which is near the null point. 
We make use of the quadrupole approximation for the magnetic field to describe this effect near the null point, $\vb{B}_{q}$ in \eqref{eq:Bq}, rewritten for the beam path as 
\begin{subequations}
\begin{align}
    B_{x} &= 0 \, , \\
    B_{y} &= \frac{2\pi B_{r}^{2}}{\mu_{0} I_{w} } z_{a} \, , \\
    B_{z} &= \frac{2\pi B_{r}^{2}}{\mu_{0} I_{w} } v \left( t- t_{\np} \right) \, .
\end{align}   
\end{subequations}
The resulting system of coupled differential equations is reduced to
\begin{subequations}\label{eq:stiff}
\begin{align}
    \label{eq:thetae} \nonumber \dot{\theta}_{e} &= -\gamma_{e} \bigl[ B_{y} \cos(\phi_{e}) \\
    &\qquad\qquad+ B_{n} \sin(\theta_{n})\sin(\phi_{n}-\phi_{e})\bigr] \ , \\
    \nonumber \dot{\phi}_{e} &= -\gamma_{e} \Bigl[ B_{z} + B_{n} \cos(\theta_{n}) - \cot(\theta_{e})\bigl[ B_{y}\sin(\phi_{e}) \\
    & \qquad \qquad + B_{n}\sin(\theta_{n})\cos(\phi_{e}-\phi_{n})  \bigr]  \Bigr] \ , \\
    \dot{\theta}_{n} &= 0 \ , \\
    \nonumber \dot{\phi}_{n} &= -\gamma_{n} \Bigl[ B_{z} + B_{e} \cos(\theta_{e}) - \cot(\theta_{n})\bigl[ B_{y}\sin(\phi_{n}) \\
    & \qquad \qquad + B_{e}\sin(\theta_{e})\cos(\phi_{e}-\phi_{n})  \bigr]  \Bigr] \ .
\end{align}  
\end{subequations}
Note that we have imposed the condition $\dot{\theta}_{n} =0$ since the postulates of the CQD theory require small variations of the polar angle of the nuclear magnetic moment.
This condition corresponds to the selection rule for nuclear spin transitions $\Delta m_{I}=0$ \cite{barh2005,wan2022}.

To solve the above differential equations, we sample the initial orientation of $\vb{\bm{\mu}}_{n}$  from \eqref{eq:hssamp} and sample $\phi_{e,0}$ uniformly, in a way similar to \eqref{eq:eiso}, while setting $\theta_{e,0} = \pi$.
The concurrence of two vastly different time scales, $\sfrac{\omega_{n}}{\omega_{e}}\approx \num{7e-5}$, for the dynamics of $\vu{\bm{\mu}}_{e}$ and $\vu{\bm{\mu}}_{n}$, leads to stiffness prone to numerical instability \cite{aik1985,haiw1996}.
We chose the Radau methods \cite{haiw1999, darks2007} implemented in Julia \cite{racn2017,bezek2017} to solve the differential equations \eqref{eq:stiff} numerically. 
Our source code is posted online  \cite{titgw2022}. 
An alternative approach, to avoid the stiffness, approximates the nuclear Larmor frequency as a constant and accelerates the solution using a variable transformation \cite{hetk2022}

\begin{figure}[t]
    \centering
    \includegraphics[width=0.97\linewidth]{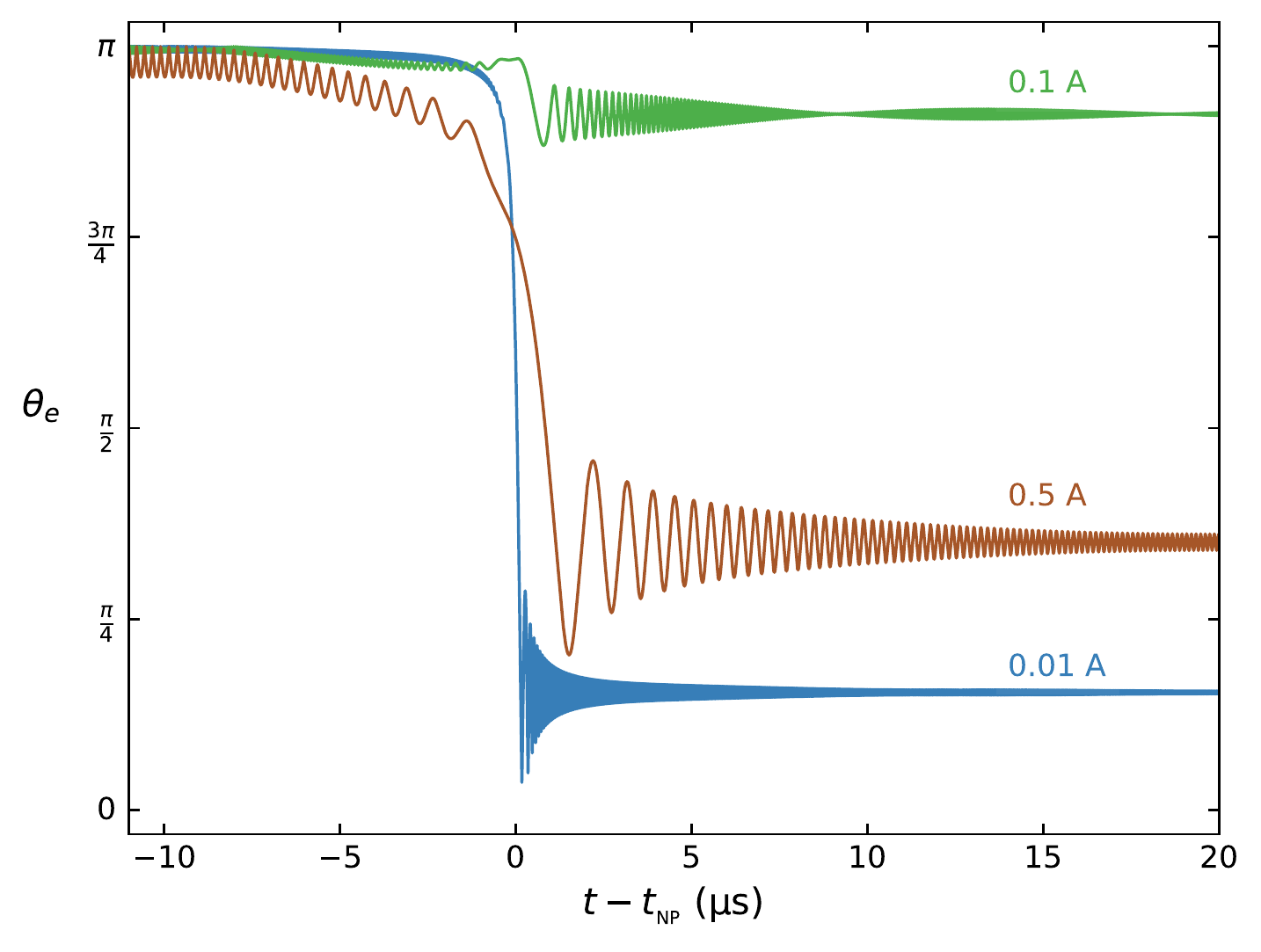}
    \caption{Dynamics of $\theta_{e}$ for different electric currents $I_{w}$. The initial conditions are $\theta_{e,0}=\pi$ and $\phi_{e,0}=0$ for the electron magnetic moment, whilst $\theta_{n,0}=\frac{5}{8}\pi$ and $\phi_{n,0}=\frac{11}{10}\pi$ for the nuclear magnetic moment.}
    \label{fig:trajectories}
\end{figure}

Figure \ref{fig:trajectories} shows the characteristic dynamics of $\theta_{e}$ inside the inner rotation chamber for a representative set of currents.
To facilitate comparison, we introduce $\Delta t = t-t_{\np}$, such that the time $\Delta t=0$ at the current-dependent null point \eqref{eq:tnp}.
The dynamics are characterized by an oscillatory behavior; the polar angle $\theta_{e}$ varies rapidly in the vicinity of the null point ($\Delta t=0$) due to strong non-adiabatic rotation.
Then, the oscillations are damped due to the increasing adiabaticity parameter for $\Delta t>0$.
However, since $\theta_{e}$ has not asymptotically converged to a definite value, we numerically approximate its final value $\theta_{e,f}$ by averaging the oscillations over the last \SI{2}{\micro\second}.

\subsubsection{Stern--Gerlach apparatus 2}

After traveling through the inner rotation chamber,  the atom enters the second Stern--Gerlach apparatus as shown in Fig.~\ref{fig:setup}.
The strong magnetic field along the $z$-axis realigns the electron magnetic moment. 
The final orientation $\theta_{e,f} \longmapsto \theta_{e,D}$ follows the branching condition postulated by the CQD theory \cite{wan2022}:
\begin{equation}
    \theta_{e,D} = 
    \begin{cases}
            0 & \text{if} \quad \theta_{e,f} < \theta_{n,0} , \\
            \pi & \text{if} \quad \theta_{e,f} > \theta_{n,0} . 
    \end{cases}
\end{equation}
Therefore, the measured polar angle $\theta_{e,D}$ takes on either $0$ or $\pi$.

\subsubsection{Fraction of spin flip}

The spin flip corresponds to those atoms with $\theta_{e,D}=\pi$, as depicted in Fig.~\ref{fig:mm}.
We numerically solved the Bloch equations for $N=\num{15000}$ atoms for each current $I_{w}$. 
Thus, the fraction of spin flip, for a given $I_{w}$, is computed using
\begin{equation}\label{eq:fsf}
    \mathcal{W}_{\mathrm{num}}= \frac{1}{N} \sum_{i=1}^{N} \left[\theta^{(i)}_{e,D}=\pi\right] , 
\end{equation}
where $i$ denotes the $i$th-atom sampled and $\left[ P \right]$ is the Iverson bracket, which is defined to take on the value 1 when the statement $P$ is true and 0 otherwise.

\begin{figure}[t]
    \centering
    \includegraphics[width=0.95\linewidth]{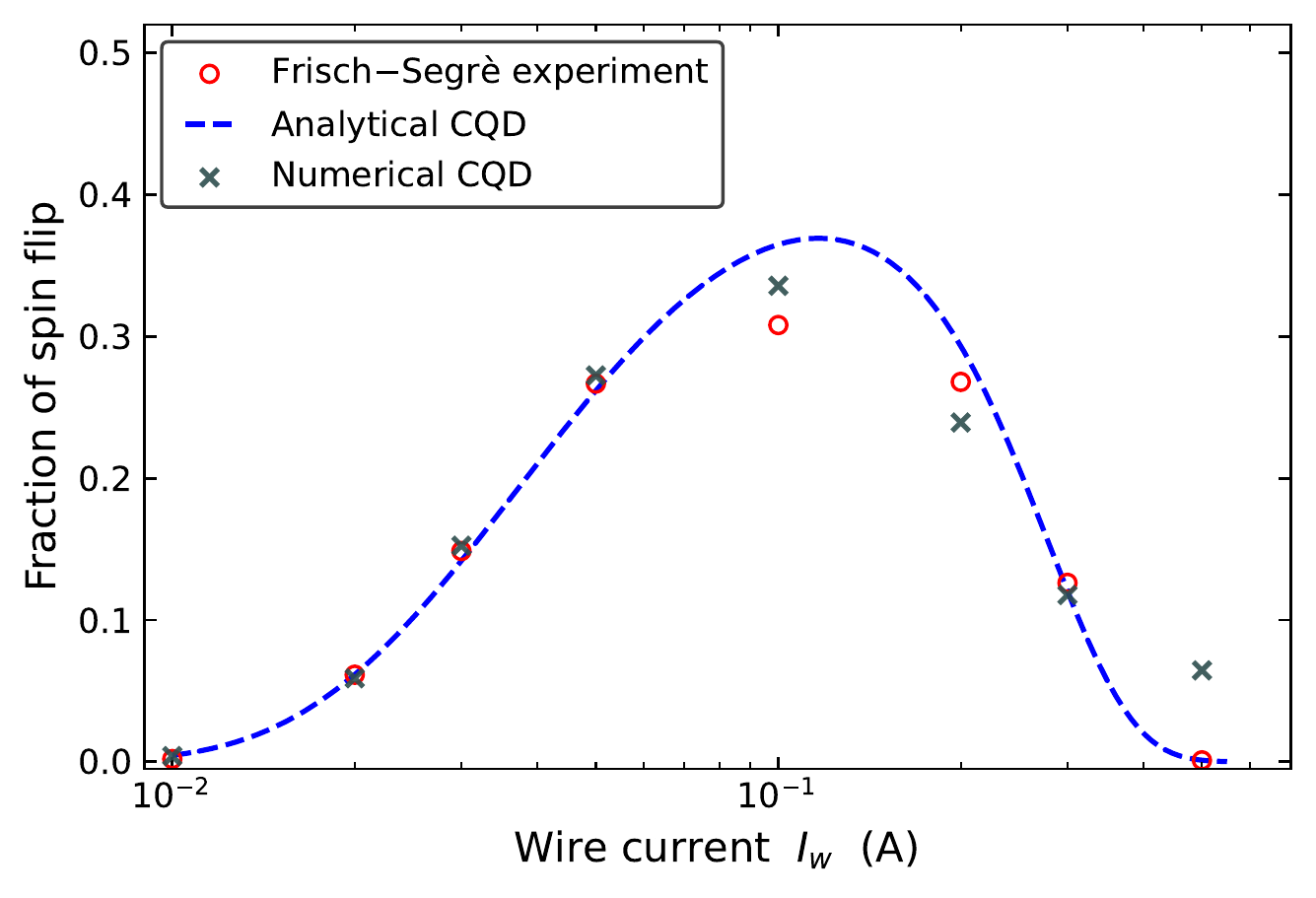}
    \caption{Fraction of spin flip for the multi-stage Stern--Gerlach experiment by Frisch and Segr\`e. The red circles are the experimental data \cite{fris1933}, the blue dashed line is the closed-form analytical solution from CQD \cite{wan2022}, and the gray crosses show the numerical solution from CQD.}
    \label{fig:result}
\end{figure}

Figure \ref{fig:result} shows the resulting fraction of spin flip obtained from the numerical simulation in comparison with the experimental results reported by Frisch and Segr\`e \cite{fris1933}.
In addition, the closed-form analytical prediction from the CQD theory is included (see Appendix \ref{app:CQD}) \cite{wan2022}.

The coefficient of determination for our numerical simulation is $R_{\mathrm{num}}^{2}=0.945$, compared with $R^{2}_{\mathrm{ana}}=0.962$ from the analytical solution \cite{wan2022}. 
The statistical errors associated with the numerical results are smaller than the symbol size in Fig.~\ref{fig:result}.
These results show a close agreement of CQD with the experimental data \cite{fris1933}, which has not yet been achieved using standard quantum mechanical treatments \cite{kahth2022}.

\section{Conclusions}\label{conclu}

We used CQD to numerically model the Frisch--Segr\`e experiment without using any fitting parameters.
The numerical problem is stiff because of the two vastly different characteristic time scales \cite{wan2022}. The solver of the differential equations must be carefully chosen and validated.
We described all the stages of the experiment, with emphasis on the inner rotation chamber.
Using the Monte Carlo method, we sampled the spatial orientations of the electron and nuclear magnetic moments of the atoms in the atomic beam.
Our findings show that the fraction of spin flip obtained by means of CQD closely reproduces the reported experimental results, with a coefficient of determination of $R^{2}_{\mathrm{num}} = 0.945$.
The obtained results have not been explained by standard quantum mechanical approaches \cite{kahth2022}, despite attempts by Majorana \cite{maj1932,kofis2022,ivasn2023} and Rabi \cite{rab1936,rab1937,blor1945}, whose theoretical formulae deviate from the experimental results.

Hence, this work supports CQD as a model for both the evolution and the collapse of the electron spin in atoms in the presence of external magnetic fields.
Some other reported experiments involving multi-stage Stern--Gerlach experiments \cite{schb1983, higrt1977} could be explored using the present formalism.
Notwithstanding, we hope that this work will encourage further experiments to verify CQD.

\begin{acknowledgments}
This project has been made possible in part by grant number 2020-225832 from the Chan Zuckerberg Initiative DAF, an advised fund of the Silicon Valley Community Foundation.
\end{acknowledgments}

\section*{Supplemental Material}
Our source code written in Julia is available online \cite{titgw2022}.

\appendix

\section{Approximation to quadrupole field}\label{app:quadru}

The total magnetic field inside the inner rotation chamber, $\vb{B} = \vb{B}_{w} + \vb{B}_{r}$, can be approximated as an ideal quadrupole field.
The Taylor expansion of $\vb{B}$ in the $yz$ plane around the null point $(y_{\np},-z_{a})$ reads 
\begin{multline}\label{eq:Tayexp}
    \vb{B} = \vb{B}\Big|_{\np} + \pdv{\vb{B}}{y} \Big|_{\np} (y-y_{\np}) +\pdv{\vb{B}}{z} \Big|_{\np} (z+z_{a}) \\
    +\dfrac{1}{2} \pdv[2]{\vb{B}}{y} \Big|_{\np} (y-y_{\np}) + \dfrac{1}{2}  \pdv[2]{\vb{B}}{z} \Big|_{\np} (z+z_{a}) \\
    +\pdv{\vb{B}}{y}{z} \Big|_{\np} (y-y_{\np})(z+z_{a}) + \cdots \, .
\end{multline}
Up to the first order, for the hyperbolic field approximation, the coefficients are
\begin{subequations} \label{eq:Taycoeff}
    \begin{align}
    \vb{B}\Big|_{\np} &= 0 \, , \\
    \pdv{\vb{B}}{y} \Big|_{\np} &= \frac{2\pi B_{r}^{2}}{\mu_{0} I_{w}} \, \vu{z} \, , \\
    \pdv{\vb{B}}{z} \Big|_{\np} &= \frac{2\pi B_{r}^{2}}{\mu_{0} I_{w}} \, \vu{y} \, .
\end{align}    
\end{subequations}

Substituting \eqref{eq:Taycoeff} into \eqref{eq:Tayexp}, the quadrupole magnetic field can be written as
\begin{equation}
    \vb{B}_{q} = \frac{2\pi B_{r}^{2}}{\mu_{0} I_{w}} (z+z_{a}) \: \vu{y} + \frac{2\pi B_{r}^{2}}{\mu_{0} I_{w}} (y-y_{\np}) \: \vu{z} .
\end{equation}
This corresponds to the magnetic field in \eqref{eq:Bq}.

\newpage
\section{Closed-form solution of the co-quantum dynamics theory}\label{app:CQD}

Using the CQD theory, it is possible to derive a closed-form formula for the probability of spin flip in the inner rotation chamber in the presence of the quadrupole magnetic field and the nuclear magnetic moment as we described for the experiment conducted by Frisch and Segr\`e \cite{wan2022}.

In terms of the parameters of the experiment, the fraction of spin flip, shown as the blue dashed line in Fig.~\ref{fig:result}, is given by
\begin{equation} \label{eq:SFcqd}
    \mathcal{W}_{\mathrm{ana}} (I_{w}) = \exp\left[ -\sqrt{ \left( \dfrac{c_{r0}}{I_{w}} \right)^{2} + c_{rs}^{2} } - c_{rr} I_{w}^3 \right] , 
\end{equation}
where
\begin{subequations}
\begin{align}
    c_{r0} &= \abs{\gamma_{e}} \dfrac{ 2\pi^{2} z_{a}^{2}}{\mu_{0} v } \left( B_{r} + B_{n}\cos(\ev{\theta_{n}}) \right)^{2} , \\
    c_{rs} &= \abs{\gamma_{e}} \dfrac{\pi z_{a}}{v} B_{n} \sin(\ev{\theta_{n}}) , \\
    c_{rr} &= \dfrac{\mu_{0}^{3} \gamma_{e}^{2}\gamma_{n}}{32\pi v^{3}} \dfrac{B_{e}\left( B_{n} \sin(\ev{\theta_{n}}) \right)^{5}}{\left( B_{r} + B_{n}\cos(\ev{\theta_{n}})  \right)^{6}} .
\end{align}
\end{subequations}
The coefficients represent three physical effects identified in the solution, \textit{i.e.,} null-point rotation, rotation saturation, and resonant rotation, respectively. 
The mean polar angle of the nuclear magnetic moment is computed from the probability density function \eqref{eq:hsd} as follows:
\begin{equation}
    \ev{\theta_{n}} = \int\dd{\theta_{n}} \dd{\phi_{n}} \sin(\theta_{n}) p_{n}(\theta_{n},\phi_{n}) = \frac{5}{8}\pi .
\end{equation}

\bibliography{ref}

\end{document}